\documentclass[showpacs,aps,floatfix,prx,reprint,superscriptaddress,nofootinbib]{revtex4-1} % STEFANO
\usepackage{amsmath}
\usepackage{graphicx}
\usepackage{color}
\usepackage{setspace}
\usepackage{multirow}
\usepackage[]{titlesec}
\usepackage{enumerate}
\usepackage{units}
\usepackage{array}
\usepackage{epstopdf}
\usepackage{float}
\usepackage{titlesec}
\usepackage[titletoc,toc,title]{appendix}
\usepackage{lipsum}
\usepackage{hyperref} \hypersetup{colorlinks=true,citecolor=blue,linkcolor=blue,urlcolor=blue} % HYPERLINKS
\usepackage{booktabs,longtable}
\definecolor{Gray}{gray}{0.9}
\definecolor{LightCyan}{rgb}{0.88,1,1}
\makeatletter \renewcommand\frontmatter@abstractwidth{\dimexpr\textwidth\relax} \makeatother  % WIDEABSTRACT
\usepackage{soul,colortbl}
\definecolor{pranab_green}{rgb}{0.31,0.53,0.10}

\def\AFLOW{{\small AFLOW}}    % AFLOW-TEX COMMON
\def\ATAT{{\small ATAT}}
\def\CALPHAD{{\small CALPHAD}}    % AFLOW-TEX COMMON
\def\GQCA{{\small GQCA}}    % AFLOW-TEX COMMON
\def\VASP{{\small VASP}}    % AFLOW-TEX COMMON
\def\MC{{\small MC}}    % AFLOW-TEX COMMON
\def\sAFLOW{{\substack{\scalebox{0.5}{AFLOW}}}}    % AFLOW-TEX COMMON
\def\sKL{{\substack{\scalebox{0.6}{KL}}}}    % AFLOW-TEX COMMON
\def\sMC{{\substack{\scalebox{0.6}{(MC)}}}}    % AFLOW-TEX COMMON
\def\smodel{{\substack{\scalebox{0.6}{(model)}}}}    % AFLOW-TEX COMMON
\def\citeAFLOW{\cite{curtarolo:art65,curtarolo:art58,curtarolo:art110,curtarolo:art63,curtarolo:art57,curtarolo:art49,curtarolo:art87,curtarolo:art121}}      % AFLOW-TEX COMMON
\def\citeAFLOWLIB{\cite{curtarolo:art75,curtarolo:art92,curtarolo:art104,curtarolo:art128}}      % AFLOW-TEX COMMON
\def\HEA{{\small HEA}}    % AFLOW-TEX COMMON
\def\HT{{\small HT}}    % AFLOW-TEX COMMON
\def\HEAs{{\small HEA}s}    % AFLOW-TEX COMMON
\def\CV{{\small CV}}    % AFLOW-TEX COMMON
\def\LTVC{{\small LTVC}}

\begin{document}
\title{\Large The search for high entropy alloys: a high-throughput {\it ab-initio} approach}

\author{Yoav Lederer}
\affiliation{Department of Mechanical Engineering and Materials Science, Duke University, Durham, North Carolina 27708, USA}
\affiliation{Department of Physics, NRCN, Beer-Sheva, 84190, Israel}
\author{Cormac Toher}
\affiliation{Department of Mechanical Engineering and Materials Science, Duke University, Durham, North Carolina 27708, USA}
\author{Kenneth S. Vecchio}
\affiliation{Department of NanoEngineering, University of California San Diego, La Jolla, CA 92093}
\author{Stefano Curtarolo}
\email{stefano@duke.edu}
\affiliation{Materials Science, Electrical Engineering, Physics and Chemistry, Duke University, Durham NC, 27708} 
\affiliation{Fritz-Haber-Institut der Max-Planck-Gesellschaft, 14195 Berlin-Dahlem, Germany}

\date{\today}

\begin{abstract}
  \noindent
  While the ongoing search to discover new high-entropy systems is slowly expanding beyond metals, a rational and effective method for predicting 
  {\it ``in silico''} the solid solution forming ability of multi-component systems remains yet to be developed.
  In this article, we propose a novel high-throughput approach, called ``\LTVC'', for estimating the transition temperature of a solid solution:  {\it ab-initio} energies are
  incorporated into a mean field statistical mechanical model where an order parameter follows the evolution of disorder.
  The \LTVC\ method is corroborated by Monte Carlo simulations and the results from the current most reliable data for binary, ternary, quaternary and quinary systems 
  (96.6\%;  90.7\%; 100\% and 100\%, of correct solid solution predictions, respectively).  
  By scanning through the many thousands of systems available in the \AFLOW\ consortium repository,
  it is possible to predict a plethora of {previously unknown potential} quaternary and quinary solid solutions for future experimental validation.
\end{abstract}

\maketitle

\section{INTRODUCTION}  \label{introduction} 

\underline{H}igh-\underline{E}ntropy \underline{A}lloys (\HEAs) are multi-component alloys forming {highly disordered} solid solution phases \cite{hea1,hea2,calph2,Gorsse_CCA_AM_2017}.
Since their discovery, just over a decade ago, \HEAs\ have attracted
the interest of the scientific community, for promising properties and
potential applications (see \cite{HEAapp1,HEAapp2,HEAprop1,HEAprop2}
as well as Ref. \onlinecite{Jien2015high} and references therein).
{The term \HEAs, and related terms such as Multiple Principle Element Alloys \cite{calph2}
and Complex Concentrated Alloys \cite{Gorsse_CCA_AM_2017} often refer to similar alloying concepts.
While there may be ongoing discussions in the literature regarding these terms, the approach outlined here is equally applicable to any of these classifications.  
For the sake of brevity, only the acronym \HEAs\ will be used throughout this article.}
The ongoing search to discover new high-entropy systems has recently expanded beyond metals to include entropy stabilized ceramics such as high-entropy oxides and carbides \cite{curtarolo:art99}.
At the time of the discovery, it was conjectured that configurational entropy was the stabilizing mechanism 
and that many multi-component alloys would form a single phase solid solution. 
However, further attempts have shown that this is valid only for a fraction of multi-component alloys, while the rest form multiple phases~\cite{hea5}.
Therefore, the key factors governing the formation of single phase \HEAs\ remain unknown \cite{Zhang2015high}.

Several semi-empirical methods have been proposed to predict which multi-component alloys will form a solid solution 
(see Ref. \onlinecite{Gao2015highdesign} for an extensive review).
Most approaches use descriptors as screening tools \cite{deshea1,deshea2,deshea3,deshea4,deshea5,deshea6,deshea7}
with parameters fitted to the available, yet limited, experimental data.
Modeling phase diagrams by using \CALPHAD\ has also been applied \cite{CoFeMnNi,calph1,calph2}, and it also suffers from insufficient experimental data.
Consequently, robust prediction of solid solution forming ability in multi-component alloys remains a major challenge hindering further \HEA\ discovery.

Phase diagram construction of multi-component alloys based on \emph{ab-initio} calculations is a direct method that can compensate for unavailable experimental data (comprehensive review by Widom \cite{widom2015high,widom1,widom2}).
Computationally very demanding, it involves energy calculations for many configurations and the implementation of statistical mechanical models for estimating thermodynamic properties \cite{deFontaine_ssp_1994,atat3}.
Hitherto, it is not surprising that the application of {\it ab-initio} searches for multi-component alloys has been considered unfeasible and without a predictive role in the search for new \HEAs.

Here, a novel \underline{h}igh-\underline{t}hroughput (\HT, \cite{curtarolo:art81}) {\it ab-initio} method is introduced 
{--- called \LTVC\ (\underline{L}ederer-\underline{T}oher-\underline{V}ecchio-\underline{C}urtarolo) ---} 
incorporating energy calculations into a mean field statistical mechanics model,
and making use of order parameters for predicting the transition temperature of a multi-component system into a solid solution phase. 
The idea is the following:
{\bf i.} The \AFLOW\ \citeAFLOW\ set of repositories \citeAFLOWLIB\ for \emph{ab-initio} calculations are leveraged to train
cluster expansion ({\small CE}) \cite{deFontaine_ssp_1994, sanchez_ducastelle_gratias:psca_1984_ce} models, within 
the \underline{A}lloy \underline{T}heoretic \underline{A}utomated \underline{T}oolkit (\ATAT) \cite{atat4} %\cite{atat1, atat2, atat3, atat4} 
and estimate zero temperature energies of atomic configurations, which are derivative structures of either fcc or bcc lattices, on which \HEAs\ show solid solution forming ability.
{\bf ii.} Next, these atomic configurations are incorporated into a mean field statistical mechanical model, named the \underline{g}eneralized \underline{q}uasi-\underline{c}hemical \underline{a}pproximation (\GQCA) \cite{GQCA1, GQCA2}, 
which is particularly suitable when long-range order is not important and the material is spatially homogeneous, as expected for solid solutions. 
{\bf iii.} Finally order parameters are proposed to detect order-disorder phase transitions by following the evolution of the statistical population of ordered configurations.

The predictive capability of {\LTVC} is corroborated by Monte Carlo simulations, experimental data for binary alloys \cite{Massalski}, \CALPHAD\ calculations performed with Thermo-Calc \cite{Andersson_CALPHAD_2001_THERMOCALC_DICRA,Thermocalc_software} 
for ternary alloys, and experimental data of 17 quaternary and quinary alloys shown by experiments to form solid solutions \cite{familyfcc,CoCrFeNi,CoFeMnNi,CoFeNiPd,AlNbTiV,HfNbTiZr,MoNbTaW,MoNbTaW2,NbTaTiV,CrNbVTiZr,MoNbTiVZr,hea2,AlCrMoTiW,HfNbTaTiZr,MoNbTaTiV,HfNbTiVZr}. 

Finally, {applying \LTVC\ to} quaternary and quinary systems, numerous alloys with solid solution forming ability are identified. 
These predictions, inaccessible by previously suggested descriptors, show that the method could become an effective guiding tool for \HEA\ design,
as well as demonstrate the importance of short-range order in these systems.

\section{METHODS}\label{methods}  
% \section*{Results}

\noindent {\bf Generation of derivative structures.} 
\HEAs\ form single-phase solid solutions mostly on fcc and bcc lattices \cite{Gao2015highdesign}: the starting point is the generation of inequivalent atomic decorations of those lattices.
A group-theoretical approach \cite{gus_enum1, gus_enum2} is used to generate a complete set of inequivalent atomic configurations with up to 8 atoms and 5 species per primitive cell (Table \ref{DS}), 
and the multiplicity	(degeneracy, number of symmetrically equivalent configurations) of each configuration is calculated.
The algorithm is validated with the binary and ternary configurations generated by the \verb|mmaps| code of the \ATAT\ package \cite{atat2}.
\begin{table}[h!]
  \centering
  \begin{tabular}{|c||c|c|c|c|} \hline 
    at./cell &  binary & ternary & quaternary & quinary \\ \hline \hline 
    % 1 & -  & - & - & - \\ \hline 
    2 & 2  & - & - & - \\ \hline 
    3 & 6 & 3 &  - & - \\ \hline 
    4 & 19 & 39 & 19 & - \\ \hline 
    5 & 28 & 81 & 108 &54\\ \hline 
    6 & 80 & 550& 1,360 & 1,500\\ \hline 
    7 & 104 & 933 & 3,876 & 7,600\\ \hline 
    8 & 390 & 6,312 & 38,372 & 111,915 \\ \hline 
  \end{tabular} 
  \caption{\small Number of inequivalent atomic configurations for fcc and bcc derivative structures including species permutations.}
  \label{DS}
\end{table}

\begin{figure*}
  \includegraphics[width=0.99\textwidth]{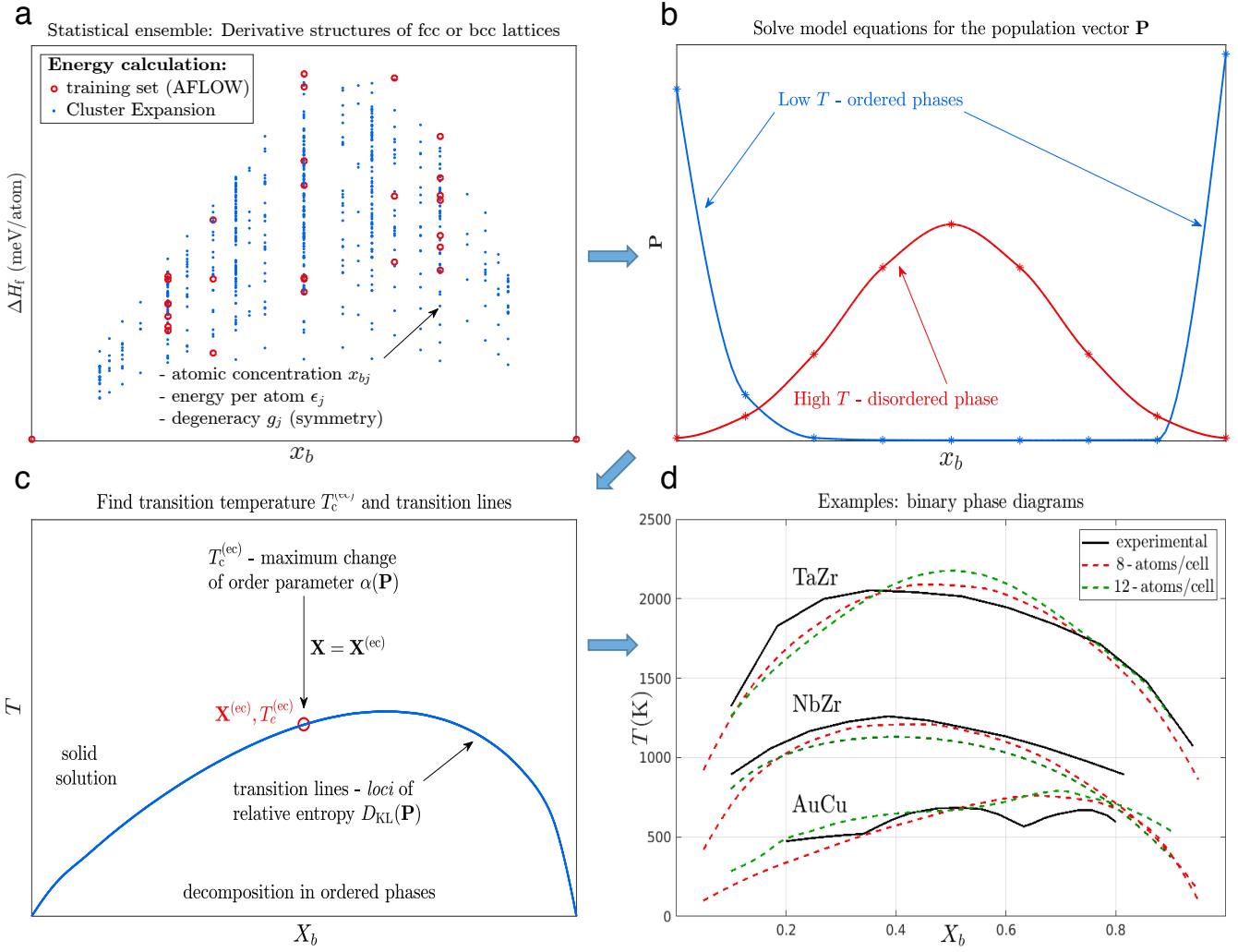}
  \vspace{-3mm}
  \caption{\small 
    {\bf Outline of the {\LTVC\ method}.}
    {\bf (a)}  Statistical ensemble construction: atomic configurations are generated, multiplicities are calculated, and energies are estimated using cluster expansion trained by the \AFLOW\ dataset.
    {\bf (b)} The statistical mechanical model is solved for the population vector ${\bf P}({\bf X},T)$ (Eq.~(\ref{population})). 
    The high-$T$ limit represents a fully disordered phase (Eq.~(\ref{po})), and the low-$T$ phase separation or ordered compound formation.
    {\bf (c)} $T_{\mathrm c}$ at equi-composition is identified with maximal change of the order parameter $\alpha$ (Eq.~(\ref{alpha})). Transition lines are traced as the {\it loci} of the relative entropy, $D_{\mathrm{KL}}$ (Eq.~(\ref{KB})).
    {\bf (d)} Examples of calculated binary phase diagrams{, using a factorization of 8-atom cells (dashed red lines) and 12-atom cells (dashed green lines). Solid black lines} represent experimental results, extracted from Ref.~\cite{Massalski}.} 
  \label{fig1}   
\end{figure*}

\noindent {\bf Calculation of zero temperature energies.} 
The \AFLOW.org repositories comprise calculated energies of relaxed atomic configurations, 
which are fcc and bcc derivative structures for most binaries and many ternaries. 
All {\it ab-initio} energies are obtained using the \VASP\ software \cite{vasp} within the \AFLOW\ high-throughput framework \citeAFLOW\ and using the standardized set of parameters \cite{curtarolo:art104}.
{The vibrational formation enthalpy is usually much smaller than the configurational contribution, especially in highly disordered systems. 
This allows us to neglect phonon spectra characterization, a daunting challenge for the millions of alloy-structures under investigation \cite{curtarolo:art96, curtarolo:art115}.}
Complete information about these calculations is included in the open access \AFLOW.org materials data repository \citeAFLOWLIB. 
For each alloy system, \AFLOW\ energies are used as input for the \verb|mmaps| code of the \ATAT\ package.
Cluster expansion is performed and energies of all configurations with up to 8 atoms per primitive cell are estimated (see binary example in Figure \ref{fig1}(a)).
In addition, \verb|mmaps| outputs the cross validation (\CV) score, which is a measure for the uncertainty of predicted energies not included in the training set (lower \CV\ score implies less uncertainty).
In this article, thermodynamic analysis is performed only for systems whose \CV\ score is less than 50 meV.

\begin{figure*}
	\includegraphics[width=0.99\textwidth]{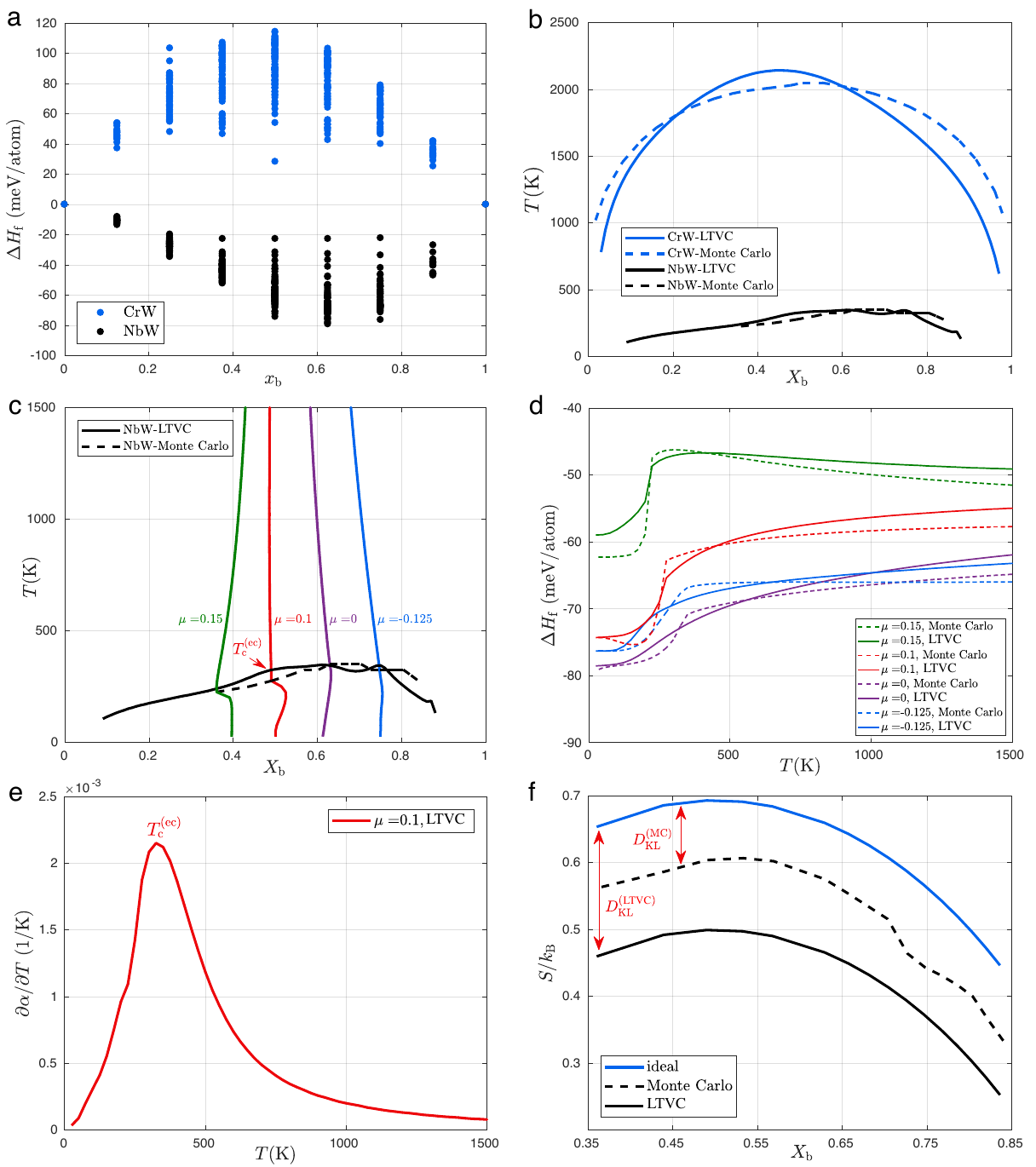}
	\vspace{-3mm}
	\caption{\small 
		{\bf  Comparison between {\LTVC} and Monte Carlo simulations.}
		{\bf (a)} Zero temperature formation enthalpies for CrW (blue) and NbW (black) from cluster expansion, which also drive the Monte Carlo calculations.  
		{\bf (b)} Transition lines for CrW (blue lines) and NbW (black lines), computed using \LTVC\ (solid lines) and Monte Carlo simulations (dashed lines).
		{\bf (c)} The NbW transition lines, with a sample of constant chemical potential trajectories, taken from the Monte Carlo simulation.
		{\bf (d)} The NbW formation enthalpies along the (${\bf X},T$), constant chemical potential  trajectories, computed using \LTVC\ (solid lines) and Monte Carlo simulations (dashed lines).
		{\bf (e)} Temperature derivative of the order parameter $\alpha$, Eq.~(\ref{alpha}), for the NbW case. The equi-composition transition temperature, $T^{\mathrm{(ec)}}_{\mathrm{c}}$, is identified as the point where
		$\partial\alpha/\partial T$ reaches its maximum value.
		{\bf (f)} Entropy, $k_{\mathrm{B}}$ normalized, across the order-disorder transition lines, computed using \LTVC\ (solid black line) and Monte Carlo simulations (dashed black lines) for the NbW case. 
		The blue line indicates the ideal mixing entropy limit. $D^{\mathrm{(model)}}_{\mathrm{KL}}$ and $D^{\mathrm{(MC)}}_{\mathrm{KL}}$ mark the relative entropy reduction, Eq.~(\ref{KB}), computed using \LTVC\ and Monte Carlo simulations, respectively. {The near constant behavior of $D^{\mathrm{(MC)}}_{\mathrm{KL}}$ supports the method's assumption.}
	} 
	\label{fig2}  
\end{figure*}

\noindent {\bf Implementation of the \GQCA\ model.} 
The atomic configurations are incorporated into a statistical mechanical mean field model, named the \underline{g}eneralized \underline{q}uasi-\underline{c}hemical \underline{a}pproximation (\GQCA) \cite{GQCA1, GQCA2}. 
This model fits well to high-throughput workflows, 
as the number of degrees of freedom varies linearly with the number of atomic types in the cell 
(compared to the cluster variation method \cite{kikuchi,deFontaine_ssp_1994}, whose number of degrees of freedom grows rapidly with cluster size).

The model factorizes a parent lattice of $N$ sites, hosting $K$ species, into non-overlapping space-filling cells of equal number of sites. 
The clusters are treated as independent, and the total energy of the $N$-site lattice is approximated as a sum of the $n$-atom cell energies, each calculated separately.
Here, the factorization is implemented using 8-atom cells, and the energy per site of each cell is estimated by assuming periodic boundary conditions on the cell surfaces:
each cell has the energy and multiplicity (symmetry degeneracy) of an analog periodic ordering, having 1, 2, 4, or 8 atoms per primitive cell.  
This implementation suits well the description of ordered or disordered phases having the same parent lattice type, as long as order effects can be captured within 8-atom cells and the spatial homogeneity assumption remains valid 
(justifying neglect of boundary interactions between phases). 
In the high-$T$ limit, a fully disordered phase (with ideal entropy of mixing) is represented by cells randomly populating the parent lattice according to the multiplicities. 
The summation of cell energies over {a} random distribution 
{to obtain} the total energy reduces a potential systematic error, related to the assumption of periodic boundary conditions. 
In the low-$T$ limit, ordered compounds or elemental phases are trivially demonstrated when only one type of unit cell (periodic ordering) occupies the whole $N$-site parent lattice.
Then, the total energy exactly becomes the sum of the cell's energies. 
In this article, the \GQCA\ model is implemented only for fcc or bcc type parent lattices. 
The incorporation of other ordered compounds, not of fcc or bcc type, into the thermodynamic analysis is described 
{below in point {\bf iv} in the section on ``Order-disorder transition''}.

Within the stated assumptions on the \GQCA\ formalism (see Refs. \onlinecite{GQCA1, GQCA2} for details), the thermodynamic potential becomes
\begin{equation} \label{FE}
  \Phi=N\cdot \left( \sum\nolimits_{j }\epsilon_jP_j-Ts-\sum\nolimits_k\mu_k\,X_k \right),
\end{equation}
where $\{P_j\}$ is the probability distribution of finding the cells with $n=8$ atoms in the $N$-site lattice, $X_k$ represents the macroscopic atomic concentration of $k$-species atoms 
on the $N$-site lattice, and $\epsilon_j$, $s$ and $\mu_k$ are the {energy of $j^{\mathrm{th}}$ cell,} configurational entropy and the chemical potentials per atom, respectively. 
The large $N$ limit for the entropy is
\begin{equation} \label{entropy}
  s=k_{\mathrm B}\left(-\sum\nolimits_kX_k\,\log(X_k)+\frac{1}{n}\sum\nolimits_jP_j\,\log(P_j/\tilde{P}_{j})\right) ,
\end{equation}
where $k_{\mathrm B}$ is the Boltzmann constant and $\{\tilde{P}_{j}\}$ is the $T$-independent random probability distribution of finding the cells in the $N$-site lattice,
\begin{equation} \label{po}
  \tilde{P_{j}}=\frac{g_j\,\prod_k X_k^{\,\,n \cdot \,x_{kj}}  }{{\sum\nolimits_{j' } g_{j'}\,\prod\nolimits_k X_k^{\,\,n\cdot\,x_{kj'}} }}.
\end{equation}
Here, $g_j$ is the multiplicity of the $j^{\mathrm{th}}$ cell and $x_{kj}$ is the fraction of $k$-species atoms.
The first term in Eq.~(\ref{entropy}) represents the ideal entropy of mixing.
The second term, 
\begin{equation} \label{KB}
  D_\sKL({\bf X},T )  \equiv\frac{1}{n}\sum\nolimits_jP_j\,\log(P_j/\tilde{P}_{j}),
\end{equation} 
is known as the relative entropy or the \emph{Kullback-Leibler} divergence \cite{RE1}, 
and is commonly used to quantify entropy loss due to ordering \cite{RE2,RE3,RE4}.
Minimizing the thermodynamic potential, subject to the constraints  
\begin{equation} \label{constraints}
  \sum\nolimits_{j}P_j=1; \quad \sum\nolimits_{j }P_j\,x_{kj}=X_k,
\end{equation}
the equilibrium probability distribution is found ($\beta=\nicefrac{1}{k_{\mathrm B} T}$):
\begin{equation} \label{population}
  P_j=\frac{\tilde{P}_{j}\,e^{-n\beta\left(\epsilon_j-\sum\nolimits_k\mu_kx_{kj}\right)}}{{\sum\nolimits_j \tilde{P}_{j}\,e^{-n\beta\left(\epsilon_j-\sum\nolimits_k\mu_kx_{kj}\right)}}}.
\end{equation}
The expression for the probability distribution is reinserted into Eq.~(\ref{constraints})
for the $(K-1)$ independent $\mu_k ({\bf X},T )$ functions of temperature and macroscopic concentration\footnote{vector ${\bf X}$ represents the macroscopic concentrations $X_k$.}.
The thermodynamic potential, Eq.~(\ref{FE}), and its related thermodynamic quantities are then easily obtained as functions of temperature and macroscopic concentration.

\noindent
{\bf Order-disorder transition.}
The random population vector $\bf \tilde{P}$, Eq.~(\ref{po}), is the high-$T$ limit of population vector $\bf{P}$, Eq.~(\ref{population}).
The limit represents a fully disordered phase with ideal entropy of mixing (only the first term in Eq.~(\ref{entropy}) being non-zero, while the relative entropy vanishes). 
An order-disorder transition {\it locus}, $\tilde{T}_{\mathrm c}({\bf X})$, can be found as the intersection between the potential of a fully disordered phase
\begin{equation} \label{Fs}
  \tilde{\Phi}_{\mathrm{disorder}} ( {\bf X},T ) \equiv \lim_{{\bf P}\to{\bf \tilde{P}}} \Phi({\bf X},T),
\end{equation}
with that of a competing ordered phase, $\Phi^\sAFLOW_{\mathrm{order}}({\bf X})$, estimated by using the \AFLOW\ convex-hull database \cite{curtarolo:art113},
\begin{equation} \label{entropic temperature}
  \tilde{\Phi}_{\mathrm{disorder}} ({\bf X},\tilde{T}_{\mathrm c}) =\Phi^\sAFLOW_{\mathrm{order}}({\bf X}) \Rightarrow \tilde{T}_{\mathrm c}({\bf X}).
\end{equation}
The approach, equivalent to a common-tangent construction of the two asymptotic free energies, is straightforward. 
Unfortunately, it disregards the effect of short range order on both the energy and entropy of the disordered phase, potentially leading to erroneous transition temperatures.

To overcome this issue,
a two-step algorithm (marked {\bf i}-{\bf ii}) is proposed.
In spirit, it is similar to what is done for phase-diagram construction using Monte-Carlo simulations, where an order parameter locating an order-disorder transition is followed by an algorithm tracing the boundary line. \\
\noindent  {\it {\bf i.} Estimation of the order-disorder transition temperature at equi-composition.} 
The change of the population vector from the high-$T$ limit (representing full disorder), to the low-$T$ limit (representing ordered compounds or elemental phases), motivates the introduction of  the order parameter{:}
\begin{equation} \label{alpha}
  \alpha\left({\bf X},T \right) \equiv {\bf P}\cdot {\bf \tilde{P}} / | {\bf P}|| {\bf \tilde{P}}|,
\end{equation}
which measures the deviation of the population vector from the high-$T$ limit.
Similar to the behavior of the order parameter in Monte-Carlo simulations \cite{atat3}, maximal change is expected at order-disorder transition {(see Figure \ref{fig2}(e)).}
The transition temperature at equi-composition $T^{\mathrm{(ec)}}_{\mathrm{c}}$ is identified as the point where
$\partial\alpha/\partial T$ reaches its maximum value (red circle in Figure \ref{fig1}(c){)}. \\

\noindent
{\it {\bf ii.} Tracing the boundary lines.}
Near the solid solution boundary lines, the entropy of the disordered phase decreases due to the growth of short range order.
The boundary lines separating solid solutions and decomposition into ordered phases (blue line in Figure \ref{fig1}(c)) are found, assuming entropy reduction is approximately constant across them
\begin{equation} \label{boundary_lines}
  D_\sKL({\bf X},T_{\mathrm{c}}) \approx D_\sKL({\bf X}^{\mathrm{(ec)}},T^{\mathrm{(ec)}}_{\mathrm{c}}).
\end{equation}
Here, $D_\sKL$ is the relative entropy, defined in Eq.~(\ref{KB}),
%{(thus the boundary lines are assumed to be the {\it loci} of $D_\sKL$)},
$T^{\mathrm{(ec)}}_{\mathrm{c}}$ is the transition temperature at equi-concentration{,} ${\bf X}^{\mathrm{(ec)}}${,} and $T_{\mathrm{c}}$ is the transition temperature at ${\bf X}$. 
The approximation works well for alloys characterized by relatively simple phase diagrams {(see Figure \ref{fig2}(f))}.  
For more complex systems, involving the formation of multiple ordered phases at low temperatures (as predicted by the \AFLOW\ convex hull), a more accurate method should involve a detailed study of the population vector ${\bf P}({\bf X},T)$ at low temperatures. 
Since \HEAs\ usually form near equi-composition, step {\bf ii} is not critical, and the results reported are based on the use of only $\alpha$ as {the} order parameter. \\
\noindent
{\bf iii.} {\it Predicting the solid solution parent lattice.}            
Steps {\bf i}-{\bf ii}  are independently implemented for fcc and bcc lattices. 
The lowest thermodynamic potential, Eq.~(\ref{FE}), will determine the structure of the solid solution.  \\
\noindent  
{\bf iv. } {\it  Effect of other competing ordered phases.}  
The formation of ordered phases, not of fcc or bcc type, might raise the order-disorder transition temperature in some parts of the phase diagram.
Using the \AFLOW\ convex-hull, the minimal thermodynamic potential of all other ordered phases $ \Phi^\sAFLOW_{\mathrm{order}}({\bf X})$ is estimated, leading to a potentially higher transition temperature:
\begin{equation} \label{new temperature}
  \Phi ({\bf X},T^{\mathrm{gs}}_{\mathrm c}) = \Phi^\sAFLOW_{\mathrm{order}}({\bf X}) \Rightarrow T^{\mathrm{gs}}_{\mathrm c}({\bf X}).
\end{equation}
If $T^{\mathrm{gs}}_{\mathrm c}$ is higher than $T_{\mathrm c}$ found by steps {\bf i}-{\bf ii}-{\bf iii},
then $T^{\mathrm{gs}}_{\mathrm c}$ is used as a better estimate for the transition temperature $T_{\mathrm c}$.

%{ Deleted: Figure \ref{fig1}(d) compares calculated with experimental boundary lines, taken from Massalski {\it et al.} \cite{Massalski}.
%The method is able to trace order-disorder transitions at equi-composition when either phase separation (NbZr and TaZr), or ordered compound formation (AuCu), occurs at low temperatures.
%Calculated boundaries for NbZr and TaZr agree better with experiments than AuCu, since the later has a miscibility gap with maximum off equi-composition. 
%A comparison between the method and Monte-Carlo simulations, further corroborating the accuracy of the method, can also be found in Appendix \ref{app-mc}.}

{
\noindent
{\bf  Test of cell size factorization convergence.}
Figure \ref{fig1}(d) compares calculated with experimental boundary lines, taken from Massalski {\it et al.} \cite{Massalski}. 
To test \LTVC's convergence with respect to cell size, calculations are performed using 8-atom cells and 12-atom cells (dashed red and green lines, respectively). 
Similarity between calculated and experimental boundaries demonstrates the method's ability to follow order-disorder transitions at equi-composition when either phase separation (NbZr and TaZr) or ordered compound formation (AuCu) occur at low temperatures. 

\noindent
{\bf Comparison of the \LTVC\ method {\it versus} Monte Carlo simulations.}
Monte Carlo (\MC) simulations are accurate --- albeit computationally demanding --- tools for calculating thermodynamic properties.
\LTVC\ and MC results are compared. 
The latter are obtained with the \verb|ATAT-memc2| code \cite{atat3}. 
Comparisons provide direct and reliable benchmarks as the two approaches are driven by the same cluster expansion. 
Two representative alloys were chosen: CrW and NbW. 
Figure \ref{fig2}(a) shows the zero temperature formation enthalpies of CrW (blue) and NbW (black), as obtained from a cluster expansion on a bcc parent lattice. 
The two alloys exhibit very different energy landscapes.
The positive formation enthalpies of CrW leads to phase separation to elemental phases at low temperatures, while the negative formation enthalpies of NbW promote ordered compounds.

Figure \ref{fig2}(b) shows the order-disorder transition lines of the two alloys, using \LTVC\ (solid lines) and \MC\ (dashed lines). 
For a wide concentration range, the agreement demonstrates the ability of the method to accurately reproduce order-disorder transitions when ordered compound formation or phase separation appear at low temperature.
% Note that $\tilde{T}_{\mathrm c}$ values for the CrW case, approximated according to Eq.~(\ref{entropic temperature}), are significantly lower (1,360K at equi-composition), demonstrating the significance of short range order effects.

Figures \ref{fig2}(c-f) depicts the NbW case. Panel (c) shows a sample of constant chemical potential trajectories, obtained from the \MC\ simulations. 
Panel (d) compares the formation enthalpy along such $(X,T)$ trajectories, using \LTVC\ (solid lines) and \MC\ simulations (dashed lines). 
The agreement demonstrates the ability of the method to accurately estimate the Gibbs free energy of the alloy both at high (disorder) and low temperatures (order).

Figure \ref{fig2}(e) shows the temperature derivative of the order parameter $\alpha$ (Eq.~(\ref{alpha})) along the equi-composition trajectory ($\mu=0.1$ red line in Figure \ref{fig2}(c)). 
The sensitivity of $\alpha$ to the underlying phase transition allows easy identification of $T^{\mathrm{(ec)}}_{\mathrm{c}}=325\mathrm{K}$, close to the \MC\ prediction (275K). 
Panel (f) depicts the entropy estimated by \LTVC\ and the \MC\ simulations along their respective boundary lines.
First, despite $D^\sMC_\sKL$ --- the entropy reduction from the ideal mixing entropy limit (blue line) --- being smaller than $D^\smodel_\sKL$, Eq.~(\ref{KB}), the model predictions do not depend on the absolute value of $D_\sKL$ at $T^{\mathrm{(ec)}}_{\mathrm{c}}$, so the difference is irrelevant. 
Second, the small variation of $D^\sMC_\sKL$ justifies the working assumption about the transition lines as the {\it loci} of relative entropy, Eq.~(\ref{boundary_lines}).
In conclusion, {\LTVC}'s combination of order parameter $\alpha$ and the relative entropy $D_\sKL$ leads to boundaries very similar to those obtained from \MC.}  

\section{{RESULTS}} \label{HT}
% \section*{Discussion}
\begin{figure*}
  \includegraphics[height=70mm,width=0.99\textwidth]{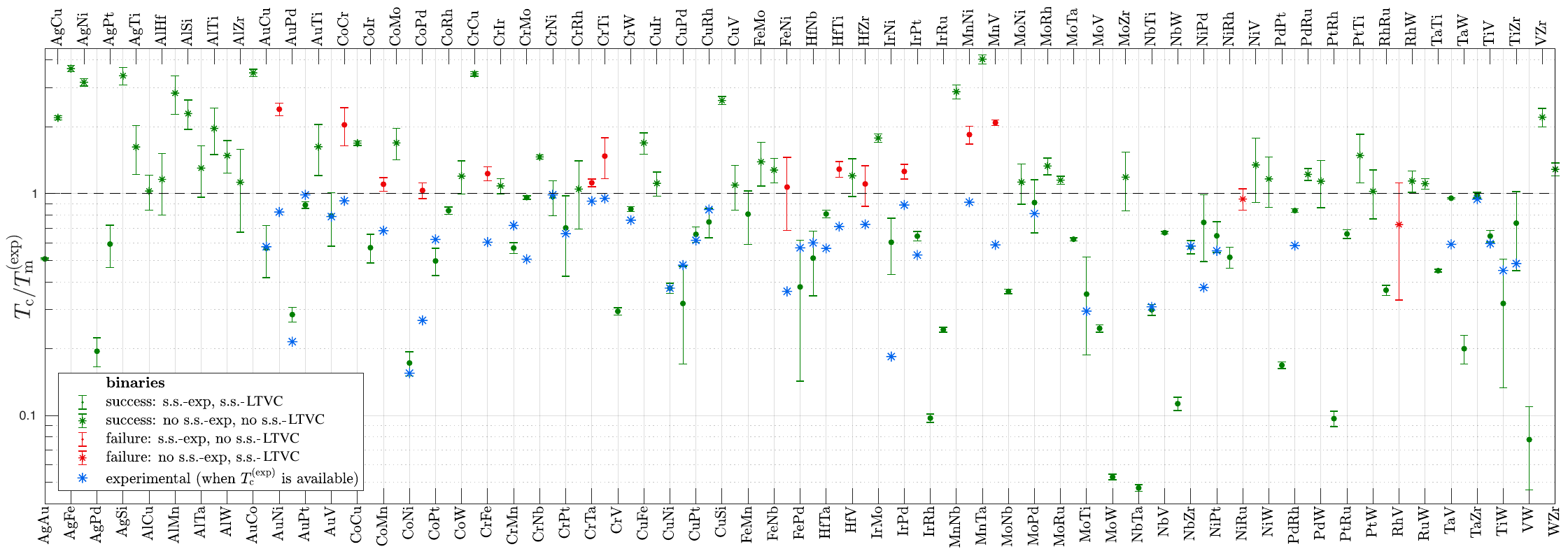}
  \caption{\small
    $T_{\mathrm c}/T^{\mathrm{(exp)}}_{\mathrm m}$ and success rate for 117 binary alloys (alphabetic
    order). A solid solution is predicted by \LTVC\
    if $T_{\mathrm c} < T^{\mathrm{(exp)}}_{\mathrm m} $ (points below the dashed line).
    {Success rate of 87.2\% is calculated as ratio of successes (in green) to total [successes+failures]; (failures in red).}  
  }
  \label{fig3}  
\end{figure*}

\begin{figure*}
  \includegraphics[width=0.99\textwidth]{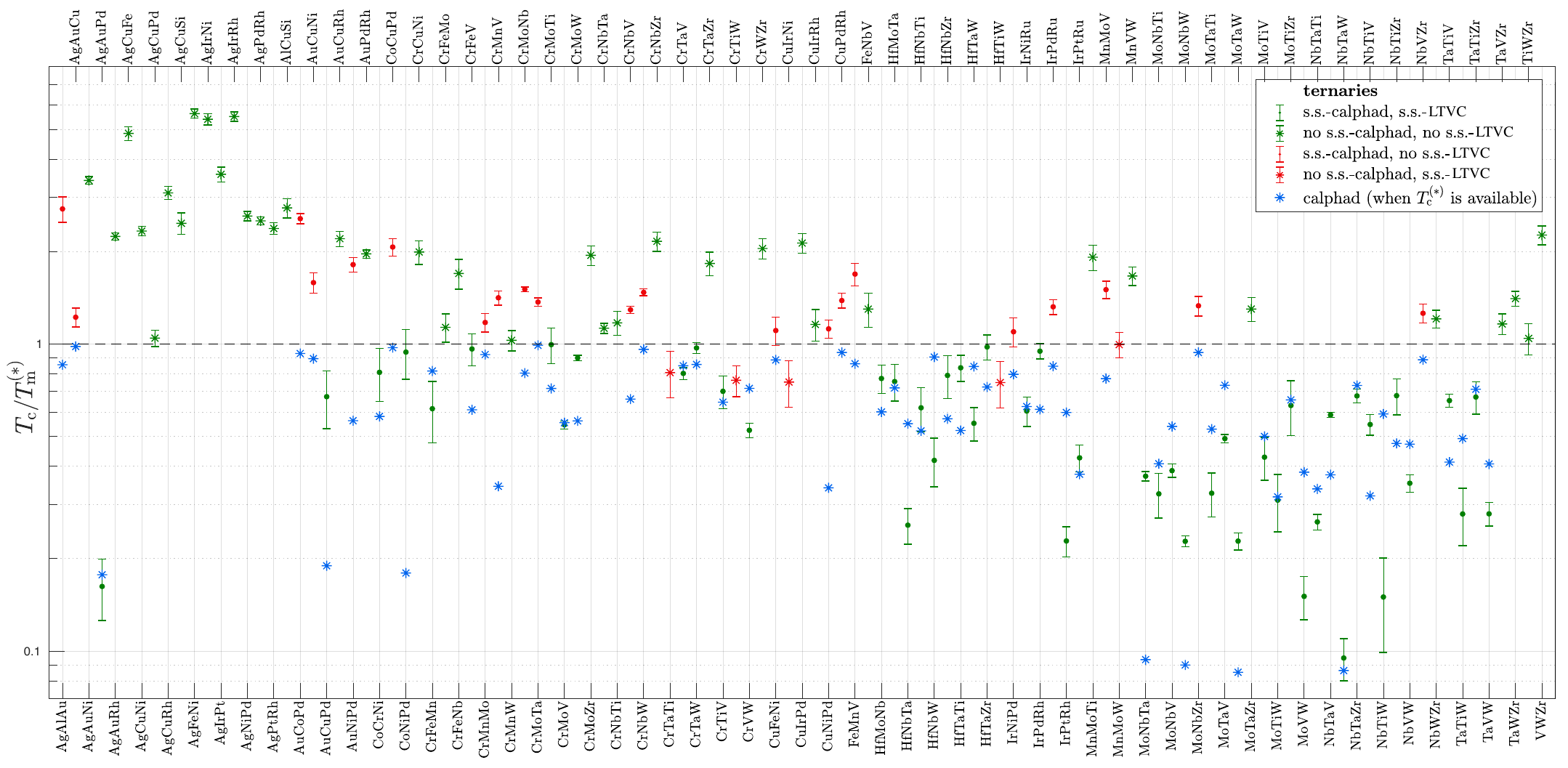}
  \vspace{-3mm}
  \caption{\small
    $T_{\mathrm c}/T^{(*)}_{\mathrm m}$ and comparison
    with \CALPHAD\  predictions for 113 ternary alloys
    (alphabetic order). $T^{(*)}_{\mathrm m}$ is the
    solidus temperature taken from \CALPHAD. A solid
    solution is predicted by \LTVC\ if $T_{\mathrm c} < T^{(*)}_{\mathrm m} $ (points below the dashed line).
    {Success rate of 77.0\% is calculated as ratio of successes (in green) to total [successes+failures]; (failures in red).}  
  }
  \label{fig4}  
\end{figure*}

\begin{figure*}
  \includegraphics[width=0.99\textwidth]{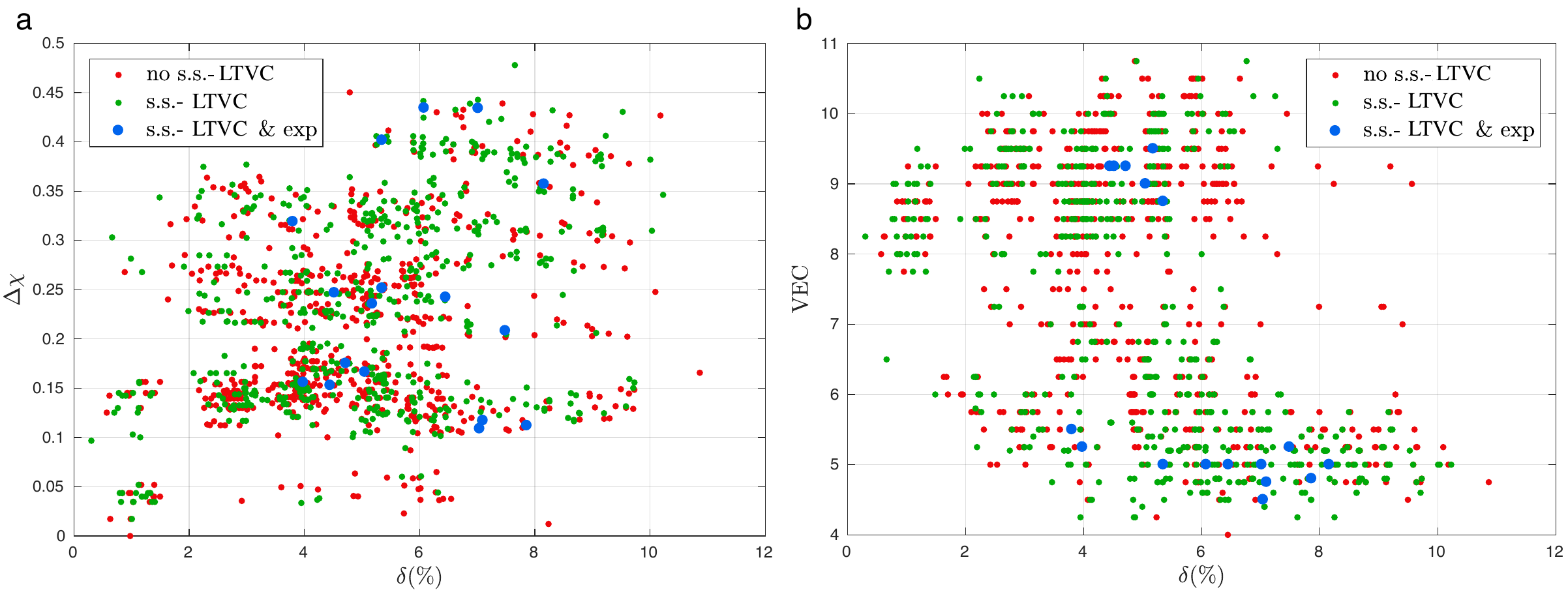}
  \vspace{-3mm}
  \caption{\small
    The failure of empirical parameters, suggested as descriptors for \HEAs, to pinpoint solid solutions. Model predictions for quaternary and quinary alloys (green - predicted s.s.,  blue - predicted s.s. and verified by experiments, see Table \ref{table-experimental}) are plotted as a function of:
    {\bf (a)}  the electro-negativity difference $(\Delta\chi)$ \cite{deshea2} and the atomic size difference $(\delta)$ \cite{deshea1};
    {\bf (b)} the valence electron concentration ({\small VEC}) \cite{deshea6} and the atomic size difference $(\delta)$.}
  \label{fig5}  
\end{figure*}

\noindent {\bf Binary alloys.}
The experimental data for 117 binary alloys is collected  \cite{Massalski} and compared with the predictions.
For each system, cluster expansion is performed separately on bcc and fcc lattices.
The energies of all inequivalent atomic configurations with up to 8 atoms per primitive cell (631 for each lattice) are estimated.
Thermodynamic analysis is then performed as described above and the transition temperature $T_{\mathrm c}$ at equi-composition is estimated. 
The standard deviation of $T_{\mathrm c}$ is estimated as follows: zero temperature energies, predicted by cluster expansion, are randomly shifted according to a normal distribution. 
The standard deviation of the normal distribution is the \CV\  score, retrieved from the output of the \verb|mmaps| code of \ATAT. 
Thermodynamic analysis is performed repeatedly, each time with new shifted energies, until convergence of the standard deviation of $T_{\mathrm c}$ is obtained.
A solid solution is predicted if $T_{\mathrm c} < T^{\mathrm{(exp)}}_{\mathrm m} $,  where $T^{\mathrm{(exp)}}_{\mathrm m} $ is the experimental melting temperature at the appropriate composition.
The results are presented in Figure \ref{fig3} and listed in Table \ref{table-binary}.
\\ $\bullet$
{\it Existence.} Formation (or not) of the solid solution is correctly reproduced in 102 out of 117 analyzed systems with an effectiveness of $\eta_{\mathrm {ss}}=102/117=87.2\%$. 
\\ $\bullet$
{\it  Solid solution forming systems.}  56 out of the 58 predicted solid solution-forming are experimentally corroborated with correct underlying lattice: $\eta_{\mathrm {latt}}=56/58=96.6\%$. 
In such cases, \LTVC\ tends to overestimate the transition temperature by $\sim$11\%.
\\ $\bullet$
{\it Non solid solution forming systems.}  46 out of 59 systems predicted not to form a solid solution phase are experimentally verified as such. Erroneous prediction of 9 out of 13 alloys is explained by the estimated standard deviation of $T_{\mathrm c}$.
\\ $\bullet$
Three binaries (AuNi, MnNi, MnV) exhibit differences greater than 3 standard deviations between calculated and experimental transition temperatures.
Test Monte Carlo simulations, performed using \ATAT\ \cite{atat3},
predict $T_{\mathrm c}$ values similar to those of \LTVC.
Likely, the problem is related to the neglect of vibrational formation entropy {coming from phonon contributions} (AuNi, one of these three alloys, is known to have large vibrational entropy of formation \cite{Zunger-AuNi}), 
or to insufficient training data: the zero temperature formation enthalpy calculations (performed on primitive cells with 4 atoms or less) could be incommensurate with larger-size order effects (\textit{e.g.} magnetic),
and thus effectively overestimate the transition temperatures.
\\ $\bullet$
The two approaches $\tilde{T}_{\mathrm c}$, Eq.~(\ref{entropic temperature}), and $T_{\mathrm c}$, steps {\bf i}-{\bf iv}, produce similar results, indicating that for binaries
the effect of short range order is limited. Later, we will show that this will not be the case for ternaries.
\\

\noindent {\bf Ternary alloys.}
Next, the complete set of 4,495 ternary alloys {that} can be formed from 31 species
is addressed\footnote{Ag, Al, As, Au, Co, Cr, Cu, Fe, Ge, Hf, Ir, Mn, Mo, Nb, Ni, Os, Pd, Pt, Re, Rh, Ru,Sb, Sc, Si, Ta, Tc, Ti, V, W, Y, and Zr.}.
In the absence of an extensive database of experimentally studied ternary alloys, the predictions of the \LTVC\ method are compared with the results of calculations
employing the \CALPHAD\ methodology \cite{CoFeMnNi,calph1,calph2}.

\noindent {\it {\bf i.} CALPHAD.} 
Phase diagram calculations are performed using the Thermo-Calc software and the {\small SSOL5} database for binary and ternary alloys \cite{Andersson_CALPHAD_2001_THERMOCALC_DICRA,Thermocalc_software}. 
The following parameters are assessed at equi-composition: the {\it solidus} temperature, $T^{(*)}_{\mathrm{m}}$, the number and types of phases at this temperature, the first reaction temperature below $T^{(*)}_{\mathrm{m}}$ and the number and types of phases below the first reaction temperature. Formation (or not) of a solid solution is then tested, along with its underlying lattice and the critical temperature, $T^{(*)}_{\mathrm{c}}$. 
Although these predictions are based mainly on extrapolation of binary data, they are expected to be reasonably reliable, as long as the associated binaries of each ternary alloy are included in the {\small SSOL5} database. Cases where the solid solution phase found by Thermo-Calc is of hcp type are not included in the comparison (as \emph{ab-initio} modeling of a solid solution phase was not performed on this lattice).

\noindent   {\bf ii.} {\it Ab-initio method}.
As for binaries, cluster expansion is performed for each ternary alloy on bcc and fcc lattices using the available training data of the \AFLOW\ repositories. Cluster expansion is performed and the energies of all inequivalent configurations (9,808 for each lattice) are estimated. Thermodynamic analysis is performed for ternaries whose \CV\  score is less than 50 meV and their transition temperature $T_{\mathrm c}$ at equi-composition is obtained, along with its estimated standard deviation. A solid solution is predicted if $T_{\mathrm c} < T_{\mathrm m}^{(\star)}$. 

Results of \emph{ab-initio} and \CALPHAD\  for 441 ternaries are listed in Table \ref{table-ternary}.
Figure \ref{fig4} shows the results for 113 ternaries whose \CV\ score is less than 30 meV and the 3 associated binaries of each ternary alloy are found in the {\small SSOL5} database.
\\ $\bullet$
{\it Existence.} Formation (or not) of the solid solution is in agreement for 87 out of 113 ternaries with an effectiveness of $\eta_{\mathrm{ss}}=87/113=77.0\%$. 
\\ $\bullet$
{\it  Solid solution forming systems.} 49 out of the 54 predicted solid solution-forming cases are corroborated by \CALPHAD\  with correct underlying lattice: $\eta_{\mathrm {latt}}=49/54=90.7\%$.
The transition temperatures of these 49 ternaries are in good agreement. 
In such cases, the method predicts transition temperatures slightly higher  ($\sim$3\%) than \CALPHAD.
Formation of solid solutions is not corroborated by \CALPHAD\  in 5 out of 54 cases. 
Here, the disagreement between the two methods can be related to the standard deviation of $T_{\mathrm c}$ as well.
Note that for 4 of these 5 cases, \CALPHAD\ exhibits the same solid solution phase below $T^{(*)}_{\mathrm{m}}$, accompanied by an additional phase.	
\\ $\bullet$
{\it  Non solid solution forming systems.} 38 out of 59 non-forming predicted cases are corroborated by \CALPHAD.
Out of the 21 cases in disagreement, 15 are not explained by the standard deviations of $T_{\mathrm c}$ 
(differences are of greater than 3 standard deviations). 
As for binaries, the problem could be related to the fact that the \LTVC\ method neglects the vibrational entropy of formation, or to insufficient training data for cluster expansion calculations.
\\ $\bullet$
$\tilde{T}_{\mathrm c}$ values (approximated according to Eq.~(\ref{entropic temperature})) are lower than $T_{\mathrm c}$ by 38.3\% and 20 of its 86 predicted solid solution-forming cases are in disagreement with \CALPHAD. This shows that the effect of short range order on transition temperature is non-negligible for ternaries. 
\ \\

\noindent {\bf Search for quaternary and quinary \HEAs.}
As before, for each alloy, cluster expansion is performed separately on bcc and fcc lattices, based on the available training data of the \AFLOW\ repositories.
The energies of all inequivalent configurations (79,185 for quaternaries, 425,219 for quinaries) are estimated. Thermodynamic analysis is performed for alloys whose \CV\  score is less than 50 meV, and the transition temperature $T_{\mathrm c}$ at equi-composition is obtained. The results for 1110 quaternary alloys and 130 quinary alloys (mainly those based on the refractory elements Cr, Hf, Mo, Nb, Ta, Ti, V, W and Zr) are listed in Tables \ref{table-quaternary} and \ref{table-quinary}, respectively. 571 of the 1240 quaternaries and quinaries show solid solution forming ability. These numbers demonstrate that among multi-component alloys, one can expect to find a plethora of solid solutions.
Table \ref{table-experimental} shows results for alloys, demonstrated by experiments to form solid solutions \cite{familyfcc,CoCrFeNi,CoFeMnNi,CoFeNiPd,AlNbTiV,HfNbTiZr,MoNbTaW,MoNbTaW2,NbTaTiV,CrNbVTiZr,MoNbTiVZr,hea2,AlCrMoTiW,HfNbTaTiZr,MoNbTaTiV,HfNbTiVZr}.
\begin{table}[h!]\label{table-ss}
  \centering
  \caption{\small 
    Comparison between {\LTVC} and experimental data for quaternaries and quinaries.
    S.S. type of single phase solid solution;
    $T_{\mathrm c}$ (K) calculated transition temperature;
    $\tilde{T}_{\mathrm c}$ (K) approximated transition temperature, estimated according to Eq.~(\ref{entropic temperature});
    $\bar{T}_{\mathrm m}$ (K) melting temperature (solidus), calculated as the average of the melting temperatures at equi-composition of the binaries associated with the alloy;
    A solid solution is predicted if $T_{\mathrm c} < \bar{T}_{\mathrm m}$;
    \CV\  (meV) cluster expansion cross validation score;
    S.S. Exp. type of single phase solid solution found in experiments. 
  } 
  \setlength\extrarowheight{2pt}
  {\small
    \begin{tabular}{|c|c|c|c|c|c|c|c|} \hline 
      \HEA\ & S.S. & $T_{\mathrm c}$   & $\tilde{T}_{\mathrm c}$   & $\bar{T}_{\mathrm m}$ & \CV\  & S.S. Exp.	\\ \hline 	\hline 	
      CoCrFeNi & fcc & 1,600  & 1,000 & 1,720 \footnote{ $T^{\mathrm{(exp)}}_{\mathrm m}$=1,695K \cite{familyfcc}} &  47 & fcc   \cite{CoCrFeNi,familyfcc}	\\ \hline
      CoCrMnNi & fcc & 1,280  & 1,020 & 1,580 & 37 &  fcc \cite{familyfcc} \\	\hline
      CoFeMnNi & fcc & 1,550  & 900 & 1,600 & 48 &  fcc  \cite{CoFeMnNi,familyfcc} \\	\hline
      AlNbTiV & bcc & 1,260  &  1,070 & 1,975 & 50 &  bcc  \cite{AlNbTiV} \\	\hline
      % CrNbTiZr & none & 3,890 & 2,200 &1,880 & 24  &  none \cite{CrNbVTiZr} \\ 	\hline
      HfNbTiZr & bcc & 970  &  840 & 2,135 & 22  &   bcc \cite{HfNbTiZr} \\	\hline
      MoNbTaW & bcc & 1,010 \footnote{Monte Carlo predicts $T_{\mathrm c}=$1,280K \cite{widom1}} & 540 & 3,065& 4 &  bcc  \cite{MoNbTaW,MoNbTaW2}	\\ \hline
      NbTaTiV & bcc & 970  & 800 & 2,345 & 6 &  bcc \cite{NbTaTiV}  \\	\hline
      NbTiVZr & bcc & 1,400 \footnote{Monte Carlo predicts $T_{\mathrm c}=$1,250K \cite{widom2015high}} & 1,110 &1,955 & 16 &  bcc  \cite{CrNbVTiZr} \\	\hline
      CoCrFeMnNi & fcc & 850 & 950 & 1,640 & 43 & fcc  \cite{hea2} \\ \hline 
      AlCrMoTiW & bcc & 1,650 & 1,000 & 2,150 & 44 & bcc  \cite{AlCrMoTiW} \\ \hline 
      HfNbTaTiZr & bcc & 1,300 & 750 & 2,280 & 17 & bcc \cite{HfNbTaTiZr} \\ \hline 
      HfNbTiVZr & bcc & 1,750 & 900 & 2,020 & 21 & bcc \cite{HfNbTiVZr} \\ \hline 
      MoNbReTaW & bcc & 2,750 & 1,100 & 3,000 & 14 & bcc \cite{MoNbTaTiV} \\ \hline 
      MoNbTaTiV & bcc & 800 & 450 & 2,470 & 8 & \cite{MoNbTaTiV}  \\ \hline 
      MoNbTaVW & bcc & 1,000 & 500 & 2,800 & 5 & bcc \cite{MoNbTaW,MoNbTaW2} \\ \hline 
      MoNbTiVZr & bcc & 1,450 & 700 & 2,120 \footnote{ $T^{\mathrm{(exp)}}_{\mathrm m}$=2,380K \cite{MoNbTiVZr}}  & 17 & bcc \cite{MoNbTiVZr} \\ \hline 
      NbReTaTiV & bcc & 950 & 800 & 2,470 & 19 & bcc \cite{MoNbTaTiV} \\ \hline 
    \end{tabular} 
  }
  \label{table-experimental} 
\end{table}
The data of all experimentally studied alloys and $T_{\mathrm c}$ values of Monte-Carlo simulations \cite{widom1,widom2015high}
corroborates the method's accuracy. $\tilde{T}_{\mathrm c}$ values (approximated according to Eq.~(\ref{entropic temperature})) are significantly lower, 
demonstrating again the importance of short range order effects.

As mentioned, several empirical parameters have been proposed as descriptors for single phase \HEAs.
Until now, they were only tested versus very limited experimental data. 
The new predictions serve as a much larger test set. 
Figure \ref{fig5} shows that three of the suggested descriptors: 
the electro-negativity difference $(\Delta\chi)$ \cite{deshea2}, 
the atomic size difference $(\delta)$ \cite{deshea1} and 
the valence electron concentration ({\small VEC}) \cite{deshea6} (see \cite{hea5} for formulas and values used in this article for these parameters),
fail to pinpoint the quaternary and quinary solid solutions predicted by \LTVC. 
Although not presented here, it is noted that another descriptor, based on enthalpies of formation calculated for
binary compounds \cite{deshea7}, also fails to identify the alloys {forming} solid solutions according to \LTVC.
These findings challenge the view that these simple descriptors can serve as an effective search tool for single phase \HEAs.

\section{CONCLUSIONS}

Robust prediction of solid solution forming ability in multi-component alloys remains a major challenge hindering the discovery of novel \HEAs. 
This article introduces a novel high-throughput method {--- called \LTVC\ ---} enabling \emph{ab-initio} searches through the vast space of possible multi-component alloys of solid solutions.
Based on the synergy of \AFLOW\ repositories, cluster expansion and a straightforward, yet accurate, mean field theory model, the approach can become an effective and efficient guiding tool for \HEA\ design.  

The accuracy is corroborated by Monte Carlo simulations,  experimental data for binaries (87.2\% agreement), \CALPHAD\ calculations for ternaries (77.0\% agr.) 
and experimental data for 17 quaternary and quinary alloys ($100\%$ agr.). 
Solid solution-forming cases are confirmed with high success rate: $96.6\%, 90.7\%, 100\%$ and $100\%$  for binaries, ternaries quaternaries and quinaries, respectively. 
The underlying lattice of the solid solution is correctly predicted as well. 
Transition temperatures, when available from experiments or \CALPHAD\ predictions, are also in good agreement:
\LTVC\ predicts transition temperatures slightly higher than the experimental values for binaries  ($\sim$11\%) and \CALPHAD\ values for ternaries ($\sim$3\%).  

Cases in disagreement with experiments or \CALPHAD, when found, are likely related to the neglect of vibrational formation entropy or to insufficient training data for cluster expansion,
pointing to future directions for improvement of the method. 
{The presented results identify potential stable solid solution candidates. 
Often, sluggish kinetics is the bottleneck in achieving equilibrium, so many transition temperatures might be quite difficult to characterize experimentally.}
%{The modeling results presented here identify candidate alloys that would be expected to achieve a single phase solid solution structure under thermal equilibrium conditions.  
%However, kinetically-limited diffusion effects can make achieving thermal equilibrium difficult in some alloy systems, so there is no  guarantee that all of the predicted transition temperatures can be experimentally validated.}  
%Therefore, while the results presented here provide guidance for potential single phase solid solution forming systems; 
%there is no guarantee that they can all be validated experimentally due to kinetic limitations.}
%In many high-melting alloys, the slow atomic diffusion at low temperatures might prevent a direct observation of either phase separation or ordering, and only the high temperature disordered solid solution behavior is observed. 
%While the model is expected to correctly predict the solid solution forming ability of such alloys, there might be a discrepancy between experiments and the model's predicted transition temperature.}

Analysis of 1110 quaternary and 130 quinary alloys show that $46\%$ of them form solid solutions, {suggesting that there are ample single phase \HEAs\ yet to be discovered.
\emph{Ab-initio} modeling of hcp solid solution phases has not been performed yet, a future avenue of investigation for the model. 
Furthermore, the extension to non-equiatomic alloys still remains a daunting task without more information on concentration ranges.}

{It is proved that short range order effects are also crucial for correct solid solution-forming predictions: 
other methods, neglecting such effects on configurational entropy, 
are likely to significantly underestimate the transition temperature to a solid solution phase --- possible erroneous prediction of solid solution-formation. 
Finally, previously suggested descriptors \cite{deshea1,deshea2,deshea6,deshea7} did not produce many of the solid solutions predicted here, 
casting doubts on the capability of such descriptors to serve as effective searching tools. 
The directions suggested in this article will facilitate the critical work of acquisition of new experimental data.}

\section*{ACKNOWLEDGMENTS}
We thank Ohad Levy, Donald Brenner, Jon-Paul Maria, Corey Oses, Matthias Scheffler, and Luca Ghiringhelli for various technical discussions.
We acknowledge support by DOD-ONR (N00014-13-1-0635, N00014-11-1-0136, N00014-15-1-2863).
S.C. acknowledges the Alexander von Humboldt Foundation for financial support.
The consortium {\sf \AFLOW.org} acknowledges Duke University -- Center for Materials Genomics --- for computational support.

\vspace{3mm}

\appendix

\newpage
\begin{widetext}
  \section{\large Binary alloys at equi-composition}\label{app-binary}
  % TABLE
  \LTcapwidth=\textwidth
  \begin{center}
    \setlength{\extrarowheight}{1.5pt}
    { \small
\begin{longtable*}{|c|c|c|c|c|c|c|c||c|c|c|c|c|c|c|c|} 
        \caption[\LTVC\ results for Binary alloys]{
          \small \LTVC\ results for binary alloys. 
          S.S. type of single solid solution phase (``none'' = not found);
          $ T_{\mathrm c}$ (K) transition temperature, as estimated by the method;
          $\tilde{T}_{\mathrm c}$ (K) approximated transition temperature, estimated according to Eq.~(\ref{entropic temperature});
          \CV\  (meV) cluster expansion cross validation score;
          S.S. Exp. type of single solid solution phase found experimentally (``none" = not found);
          $T^{\mathrm{(exp)}}_{\mathrm c}$ (K) experimental transition temperature;
          $T^{\mathrm{(exp)}}_{\mathrm m}$ (K) experimental melting temperature ({\it solidus}).
        } \label{table-binary} \\
        
        \hline
        \multicolumn{1}{|c|}{} & \multicolumn{1}{c|}{S.S.} & \multicolumn{1}{c|}{$T_{\mathrm c}$}& \multicolumn{1}{c|}{$\tilde{T}_{\mathrm c}$} & \multicolumn{1}{c|}{\CV} & \multicolumn{1}{c|}{S.S. Exp.} & \multicolumn{1}{c|}{$T^{\mathrm{(exp)}}_{\mathrm c}$} &\multicolumn{1}{c||}{$T^{\mathrm{(exp)}}_{\mathrm m}$} &  
        \multicolumn{1}{c|}{} & \multicolumn{1}{c|}{S.S.} & \multicolumn{1}{c|}{$T_{\mathrm c}$}& \multicolumn{1}{c|}{$\tilde{T}_{\mathrm c}$} & \multicolumn{1}{c|}{\CV} & \multicolumn{1}{c|}{S.S. Exp.} & \multicolumn{1}{c|}{$T^{\mathrm{(exp)}}_{\mathrm c}$} &\multicolumn{1}{c|}{$T^{\mathrm{(exp)}}_{\mathrm m}$} 
        \\ 
        \endfirsthead
        
        \hline
        \multicolumn{1}{|c|}{} & \multicolumn{1}{c|}{S.S.} & \multicolumn{1}{c|}{$T_{\mathrm c}$}& \multicolumn{1}{c|}{$\tilde{T}_{\mathrm c}$} & \multicolumn{1}{c|}{\CV} & \multicolumn{1}{c|}{S.S. Exp.} & \multicolumn{1}{c|}{$T^{\mathrm{(exp)}}_{\mathrm c}$} &\multicolumn{1}{c||}{$T^{\mathrm{(exp)}}_{\mathrm m}$} &  
        \multicolumn{1}{c|}{} & \multicolumn{1}{c|}{S.S.} & \multicolumn{1}{c|}{$T_{\mathrm c}$}& \multicolumn{1}{c|}{$\tilde{T}_{\mathrm c}$} & \multicolumn{1}{c|}{\CV} & \multicolumn{1}{c|}{S.S. Exp.} & \multicolumn{1}{c|}{$T^{\mathrm{(exp)}}_{\mathrm c}$} &\multicolumn{1}{c|}{$T^{\mathrm{(exp)}}_{\mathrm m}$} 
        \\ \hline 
        \endhead 
        \hline
        
        AgAu & fcc & 660 & 750 & 1 & fcc &  & 1,300 & AgCu & none & 2,350 & 1,450 & 8 & none &  & 1,070 \\ \hline 
        AgFe & none & 6,590 & 5,400 & 22 & none &  & 1,800 & AgNi & none & 5,620 & 3,480 & 21 & none &  & 1,770 \\ \hline 
        AgPd & fcc & 300 & 310 & 7 & fcc &  & 1,540 & AgPt & fcc & 1,090 & 650 & 23 & fcc &  & 1,840 \\ \hline 
        AgSi & none & 5,270 & 5,390 & 36 & none &  & 1,550 & AgTi & none & 2,790 & 3,050 & 45 & none &  & 1,720 \\ \hline 
        AlCu & none & 1,130 & 1,480 & 21 & none &  & 1,100 & AlHf & none & 2,400 & 3,790 & 47 & none &  & 2,070 \\ \hline 
        AlMn & none & 4,170 & 2,750 & 50 & none &  & 1,470 & AlSi & none & 3,030 & 4,280 & 35 & none &  & 1,320 \\ \hline 
        AlTa & none & 2,960 & 2,540 & 48 & none &  & 2,270 & AlTi & none & 3,480 & 2,850 & 50 & none &  & 1,770 \\ \hline 
        AlW & none & 2,850 & 3,390 & 36 & none &  & 1,920 & AlZr & none & 1,970 & 3,870 & 49 & none &  & 1,750 \\
\hline 
      \end{longtable*}
    } % \small
  \end{center}

  % \vspace{6mm}
  % \newpage
  \section{\large Ternary alloys at equi-composition} \label{app-ternary}
  \vspace{-3mm}
  \begin{center}
    \setlength{\extrarowheight}{1.5pt}
    {\small
      \begin{longtable*}{|c|c|c|c|c|c|c|c|c||c|c|c|c|c|c|c|c|c|} 
        \caption[\LTVC\ results for Ternary alloys]{\small \LTVC\ results for ternary alloys.
          S.S. type of single solid solution phase (``none'' = not found);
          $ T_{\mathrm c}$ (K) transition temperature, as estimated by the method;
          $\tilde{T}_{\mathrm c}$ (K) approximated transition temperature, estimated according to Eq.~(\ref{entropic temperature});
          \CV\  (meV) cluster expansion cross validation score;
          S.S.$^{(\star)}$  type of single solid solution phase using \CALPHAD\  (``none" = not found);
          $T^{(\star)}_{\mathrm c}$ (K) transition temperature, as estimated by \CALPHAD;
          $T^{(\star)}_{\mathrm m}$ (K) melting temperature (solidus), as estimated by \CALPHAD;
          $n^{(\star)}$ number of binary alloys (3 possible for each ternary) found in the {\small SSOL5} database of Thermo-Calc.
        } \label{table-ternary} \\
        
        %	\hline System & S.S. & T_{\mathrm c}  & $\tilde{T}_{\mathrm c}$  & \CV\  & $\Delta H_{\mathrm f} $ & $\delta$   \\ \hline 
        % \endfirsthead
        
        \hline
        \multicolumn{1}{|c|}{} & \multicolumn{1}{c|}{S.S.} & \multicolumn{1}{c|}{$T_{\mathrm c}$}& \multicolumn{1}{c|}{$\tilde{T}_{\mathrm c}$} & \multicolumn{1}{c|}{\CV} & \multicolumn{1}{c|}{S.S.$^{(\star)}$} & \multicolumn{1}{c|}{$T^{(\star)}_{\mathrm c}$} &\multicolumn{1}{c|}{$T^{(\star)}_{\mathrm m}$} &  \multicolumn{1}{c||}{$n^{(\star)}$} &
        \multicolumn{1}{c|}{} & \multicolumn{1}{c|}{S.S.} & \multicolumn{1}{c|}{$T_{\mathrm c}$}& \multicolumn{1}{c|}{$\tilde{T}_{\mathrm c}$} & \multicolumn{1}{c|}{\CV} & \multicolumn{1}{c|}{S.S.$^{(\star)}$} & \multicolumn{1}{c|}{$T^{(\star)}_{\mathrm c}$} &\multicolumn{1}{c|}{$T^{(\star)}_{\mathrm m}$} &  \multicolumn{1}{c|}{$n^{(\star)}$} 
        \\ \hline 
        \endfirsthead
        
        \hline
        \multicolumn{1}{|c|}{} & \multicolumn{1}{c|}{S.S.} & \multicolumn{1}{c|}{$T_{\mathrm c}$}& \multicolumn{1}{c|}{$\tilde{T}_{\mathrm c}$} & \multicolumn{1}{c|}{\CV} & \multicolumn{1}{c|}{S.S.$^{(\star)}$} & \multicolumn{1}{c|}{$T^{(\star)}_{\mathrm c}$} &\multicolumn{1}{c|}{$T^{(\star)}_{\mathrm m}$} &  \multicolumn{1}{c||}{$n^{(\star)}$} &
        \multicolumn{1}{c|}{} & \multicolumn{1}{c|}{S.S.} & \multicolumn{1}{c|}{$T_{\mathrm c}$}& \multicolumn{1}{c|}{$\tilde{T}_{\mathrm c}$} & \multicolumn{1}{c|}{\CV} & \multicolumn{1}{c|}{S.S.$^{(\star)}$} & \multicolumn{1}{c|}{$T^{(\star)}_{\mathrm c}$} &\multicolumn{1}{c|}{$T^{(\star)}_{\mathrm m}$} &  \multicolumn{1}{c|}{$n^{(\star)}$} 
        \\ \hline 
        \endhead 

        AgAlAu & none & 2,670 & 2,530 & 27 & fcc & 830 & 970 & 3 & AgAlFe & none & 7,000 & 7,000 & 38 & none &  & 1,050 & 3 \\ \hline 
        AgAlNi & none & 6,020 & 3,690 & 37 & none &  & 1,220 & 3 & AgAlV & none & 5,160 & 2,650 & 45 & none &  & 1,040 & 3 \\ \hline 
        AgAsAu & none & 2,780 & 1,820 & 36 & none &  & 900 & 2 & AgAsCu & none & 1,920 & 1,780 & 27 & none &  & 800 & 1 \\ \hline 
        AgAsPd & none & 1,660 & 1,620 & 37 & none &  & 1,010 & 1 & AgAuCo & none & 4,880 & 2,920 & 35 & none &  & 1,310 & 2 \\ \hline 
        AgAuCu & none & 1,320 & 720 & 10 & fcc & 1,060 & 1,080 & 3 & AgAuFe & none & 6,490 & 2,680 & 41 & none &  & 1,280 & 2 \\ \hline 
        AgAuMo & none & 4,200 & 3,870 & 39 & none &  & 1,280 & 2 & AgAuNi & none & 4,840 & 1,850 & 15 & none &  & 1,420 & 3 \\ \hline 
        AgAuPd & fcc & 250 & 390 & 6 & fcc & 273 & 1,540 & 3 & AgAuPt & fcc & 1,110 & 560 & 7 & fcc & 1,070 & 1,450 & 2 \\ \hline 
        AgAuRh & none & 4,860 & 1,970 & 14 & none &  & 2,170 & 3 & AgAuSi & none & 3,100 & 2,450 & 39 & none &  & 920 & 3 \\
\hline 
      \end{longtable*}
    } %\small
  \end{center}

  \newpage
  \section{\large Quaternary alloys at equi-composition} \label{app-quaternary}
  \begin{center}
    \setlength{\extrarowheight}{1.5pt}
    {\small
      \begin{longtable*}{|c|c|c|c|c|c||c|c|c|c|c|c||c|c|c|c|c|c|} 
        \caption[\LTVC\ results for Quaternary alloys]{\small \LTVC\ results for quaternary alloys.
          S.S. type of single solid solution phase (``none'' = not found);
          $ T_{\mathrm c}$ (K) transition temperature, as estimated by the method;
          $\tilde{T}_{\mathrm c}$ (K) approximated transition temperature, estimated according to Eq.~(\ref{entropic temperature});
          \CV\  (meV) cluster expansion cross validation score;		
          $\bar{T}_{\mathrm m}$ (K) melting temperature (solidus), calculated as the average of the melting temperatures at equi-composition of the 6 binary alloys associated with the quaternary alloy;
          A solid solution is predicted if $T_{\mathrm c} < \bar{T}_{\mathrm m}$.
        } \label{table-quaternary} \\
        
        %	\hline System & S.S. & T_{\mathrm c}  & $\tilde{T}_{\mathrm c}$  & \CV\  & $\Delta H_{\mathrm f} $ & $\delta$   \\ \hline 
        % \endfirsthead
        
        \hline
        \multicolumn{1}{|c|}{} & \multicolumn{1}{c|}{S.S.} & \multicolumn{1}{c|}{$T_{\mathrm c}$}& \multicolumn{1}{c|}{$\tilde{T}_{\mathrm c}$}  & \multicolumn{1}{c|}{$\bar{T}_{\mathrm m}$} & \multicolumn{1}{c||}{\CV}  &
        \multicolumn{1}{c|}{} & \multicolumn{1}{c|}{S.S.} & \multicolumn{1}{c|}{$T_{\mathrm c}$}& \multicolumn{1}{c|}{$\tilde{T}_{\mathrm c}$} & \multicolumn{1}{c|}{$\bar{T}_{\mathrm m}$} & \multicolumn{1}{c||}{\CV}   &
        \multicolumn{1}{c|}{} & \multicolumn{1}{c|}{S.S.} & \multicolumn{1}{c|}{$T_{\mathrm c}$}& \multicolumn{1}{c|}{$\tilde{T}_{\mathrm c}$}  & \multicolumn{1}{c|}{$\bar{T}_{\mathrm m}$} & \multicolumn{1}{c|}{\CV} 
        \\ \hline 
        \endfirsthead
        
        \hline
        \multicolumn{1}{|c|}{} & \multicolumn{1}{c|}{S.S.} & \multicolumn{1}{c|}{$T_{\mathrm c}$}& \multicolumn{1}{c|}{$\tilde{T}_{\mathrm c}$}  & \multicolumn{1}{c|}{$\bar{T}_{\mathrm m}$} & \multicolumn{1}{c||}{\CV}  &
        \multicolumn{1}{c|}{} & \multicolumn{1}{c|}{S.S.} & \multicolumn{1}{c|}{$T_{\mathrm c}$}& \multicolumn{1}{c|}{$\tilde{T}_{\mathrm c}$} & \multicolumn{1}{c|}{$\bar{T}_{\mathrm m}$} & \multicolumn{1}{c||}{\CV}   &
        \multicolumn{1}{c|}{} & \multicolumn{1}{c|}{S.S.} & \multicolumn{1}{c|}{$T_{\mathrm c}$}& \multicolumn{1}{c|}{$\tilde{T}_{\mathrm c}$}  & \multicolumn{1}{c|}{$\bar{T}_{\mathrm m}$} & \multicolumn{1}{c|}{\CV} 
        \\ \hline 
        \endhead 

        \hline 
        AgAlAuCo & none & 5,000 & 5,000 & 1,320 & 36 & AgAlAuCu & fcc & 980 & 1,120 & 1,220 & 24 & AgAlAuFe & none & 3,620 & 4,360 & 1,330 & 41 \\ \hline 
        AgAlAuNi & none & 5,000 & 5,000 & 1,310 & 32 & AgAlAuSi & none & 2,640 & 3,040 & 1,300 & 29 & AgAlAuZr & none & 2,700 & 1,840 & 1,410 & 50 \\ \hline 
        AgAlCoCu & none & 5,000 & 5,000 & 1,320 & 34 & AgAlCoNi & none & 5,000 & 5,000 & 1,420 & 42 & AgAlCuFe & none & 3,160 & 1,980 & 1,330 & 32 \\ \hline 
        AgAlCuMn & none & 3,400 & 1,760 & 1,260 & 50 & AgAlCuNi & none & 5,000 & 5,000 & 1,310 & 28 & AgAlCuSi & none & 3,340 & 3,540 & 1,300 & 19 \\ \hline 
        AgAlCuZr & none & 2,260 & 2,440 & 1,410 & 48 & AgAlFeNi & none & 5,000 & 5,000 & 1,430 & 46 & AgAlNiSi & none & 1,780 & 3,220 & 1,390 & 46 \\ \hline 
        AgAuCoCr & none & 5,000 & 5,000 & 1,620 & 50 & AgAuCoCu & none & 5,000 & 5,000 & 1,320 & 19 & AgAuCoIr & fcc & 340 & 2,640 & 1,760 & 35 \\ \hline 
        AgAuCoMo & none & 5,000 & 5,000 & 1,810 & 43 & AgAuCoNi & fcc & 420 & 2,140 & 1,470 & 27 & AgAuCoPd & none & 5,000 & 5,000 & 1,460 & 16 \\ \hline 
        AgAuCoPt & none & 1,640 & 1,800 & 1,520 & 22 & AgAuCoRe & none & 5,000 & 5,000 & 1,950 & 50 & AgAuCoRh & fcc & 1,080 & 2,040 & 1,640 & 20 \\
\hline 
      \end{longtable*}
    } %\small
  \end{center}

  \section{\large Quinary alloys at equi-composition} \label{app-quinary}

  \begin{center}
    \setlength{\extrarowheight}{1.5pt}
    {\small
      \begin{longtable*}{|c|c|c|c|c|c||c|c|c|c|c|c|} 
        \caption[\LTVC\ results for Quinary alloys]{\small \LTVC\ results for quinary alloys.
          S.S. type of single solid solution phase (``none'' = not found);
          $ T_{\mathrm c}$ (K) transition temperature, as estimated by the method;
          $\tilde{T}_{\mathrm c}$ (K) approximated transition temperature, estimated according to Eq.~(\ref{entropic temperature});
          \CV\  (meV) cluster expansion cross validation score;		
          $\bar{T}_{\mathrm m}$ (K) melting temperature (solidus), calculated as the average of the melting temperatures at equi-composition of the 10 binary alloys associated with the quinary alloy;
          A solid solution is predicted if $T_{\mathrm c} < \bar{T}_{\mathrm m}$.
        } \label{table-quinary} \\
        
        %	\hline System & S.S. & T_{\mathrm c}  & $\tilde{T}_{\mathrm c}$  & \CV\  & $\Delta H_{\mathrm f} $ & $\delta$   \\ \hline 
        % \endfirsthead
        
        \hline
        \multicolumn{1}{|c|}{} & \multicolumn{1}{c|}{S.S.} & \multicolumn{1}{c|}{$T_{\mathrm c}$}& \multicolumn{1}{c|}{$\tilde{T}_{\mathrm c}$}  & \multicolumn{1}{c|}{$\bar{T}_{\mathrm m}$} & \multicolumn{1}{c||}{\CV}  &
        \multicolumn{1}{c|}{} & \multicolumn{1}{c|}{S.S.} & \multicolumn{1}{c|}{$T_{\mathrm c}$}& \multicolumn{1}{c|}{$\tilde{T}_{\mathrm c}$}  & \multicolumn{1}{c|}{$\bar{T}_{\mathrm m}$} & \multicolumn{1}{c|}{\CV} 
        \\ \hline 
        \endfirsthead
        
        \hline
        \multicolumn{1}{|c|}{} & \multicolumn{1}{c|}{S.S.} & \multicolumn{1}{c|}{$T_{\mathrm c}$}& \multicolumn{1}{c|}{$\tilde{T}_{\mathrm c}$}  & \multicolumn{1}{c|}{$\bar{T}_{\mathrm m}$} & \multicolumn{1}{c||}{\CV}  &
        \multicolumn{1}{c|}{} & \multicolumn{1}{c|}{S.S.} & \multicolumn{1}{c|}{$T_{\mathrm c}$}& \multicolumn{1}{c|}{$\tilde{T}_{\mathrm c}$}  & \multicolumn{1}{c|}{$\bar{T}_{\mathrm m}$} & \multicolumn{1}{c|}{\CV} 
        \\ \hline 
        \endhead 
        \hline 

        \hline 
        AlCrMoTiW & bcc & 1,650 & 1,000 & 2,150 & 44 & CoCrFeMnNi & fcc & 850 & 950 & 1,640 & 43 \\ \hline 
        CrHfMoNbTa & bcc & 1,900 & 1,250 & 2,440 & 19 & CrHfMoNbTi & bcc & 1,700 & 1,100 & 2,180 & 23 \\ \hline 
        CrHfMoNbV & none & 2,550 & 1,150 & 2,230 & 21 & CrHfMoNbW & bcc & 2,100 & 1,000 & 2,520 & 19 \\ \hline 
        CrHfMoNbZr & bcc & 1,800 & 1,450 & 2,150 & 26 & CrHfMoTaTi & bcc & 1,650 & 1,250 & 2,260 & 23 \\ \hline 
        CrHfMoTaV & none & 2,300 & 1,150 & 2,290 & 22 & CrHfMoTaW & bcc & 1,900 & 1,100 & 2,620 & 19 \\ \hline 
        CrHfMoTaZr & bcc & 2,050 & 1,600 & 2,200 & 26 & CrHfMoTiV & none & 2,200 & 1,100 & 2,060 & 26 \\ \hline 
        CrHfMoTiW & bcc & 1,750 & 1,000 & 2,320 & 24 & CrHfMoTiZr & bcc & 1,750 & 1,350 & 1,990 & 30 \\ \hline 
        CrHfMoVW & bcc & 2,300 & 1,150 & 2,390 & 22 & CrHfMoVZr & none & 3,000 & 1,450 & 2,000 & 30 \\
\hline 
      \end{longtable*}
    } %\small
  \end{center}
  % \end{appendices}

\end{widetext}

\end{document}